\documentclass[useAMS,usenatbib]{mn2e}
\usepackage{amsmath}
\usepackage{graphicx}

\title[Transit Timing Variations Of Circumbinary Exoplanets]{Placing Limits on the Transit Timing Variations of Circumbinary Exoplanets}

\author[Armstrong et. al.]
{\parbox{\textwidth}{D. Armstrong$^1$\thanks{E-mail: d.j.armstrong@warwick.ac.uk}, D. V. Martin$^2$, G. Brown$^1$, F. Faedi$^1$, Y. G\'{o}mez Maqueo Chew$^{1,3}$, R. Mardling$^{4,2}$, D. Pollacco$^1$, A. H.M.J. Triaud$^{5,2}$\thanks{Fellow of the Swiss National Science Foundation}, S. Udry$^2$}
\vspace{0.4cm}\\
\parbox{\textwidth}{$^{1}$University of Warwick, Department of Physics, Gibbet Hill Road, Coventry, CV4 7AL, UK\\
$^{2}$Observatoire de Gen\`eve, Universit\'e de Gen\`eve, 51 chemin des Maillettes, Sauverny 1290, Switzerland\\
$^{3}$Department of Physics \& Astronomy, Vanderbilt University, Nashville, TN 37235, USA\\
$^{4}$School of Mathematical Sciences, Monash University, Vic 3800, Australia\\
$^{5}$Department of Physics, and Kavli Institute for Astrophysics and Space Research, Massachusetts Institute of Technology, Cambridge, MA 02139, USA}}

\begin{document}

\date{Accepted 2013 July 1. Received 2013 June 28; in original form 2013 May 15}

\pagerange{\pageref{firstpage}--\pageref{lastpage}} \pubyear{2002}

\maketitle

\label{firstpage}

\begin{abstract}
We present an efficient analytical method to predict the maximum transit timing variations of a circumbinary exoplanet, given some basic parameters of the host binary. We derive an analytical model giving limits on the potential location of transits for coplanar planets orbiting eclipsing binaries, then test it against numerical N-body simulations of a distribution of binaries and planets. We also show the application of the analytic model to Kepler-16b, -34b and -35b. The resulting method is fast, efficient and is accurate to approximately 1\% in predicting limits on possible times of transits over a three-year observing campaign. The model can easily be used to, for example, place constraints on transit timing while performing circumbinary planet searches on large datasets. It is adaptable to use in situations where some or many of the planet and binary parameters are unknown.
\end{abstract}


\section{Introduction}
\label{sectintro}
To date seven transiting circumbinary exoplanets have been discovered, all from the NASA Kepler mission data \citep{Doyle:2011ev,Welsh:2012kl,Schwamb:2012ts,Orosz:2012ip,Orosz:2012ku}. Such circumbinary planets provide interesting tests of planet formation theories, having formed in a complex environment. Recently several studies have been performed on their formation \citep[e.g.][]{Gong:2012il,Pelupessy:2012gl,Meschiari:2012ts,Meschiari:2012er}, orbital stability \citep[e.g.][]{Jaime:2012br,Doolin:2011ib,Pichardo:2008et,Pichardo:2005fb} and variations in insolation from a habitability perspective \citep[e.g.][]{Kane:2012fl,OMalleyJames:2012jf}. They are beginning to be subjected to analytical models, such as that provided by \citet{Leung:2013fm} (hereafter LL). Such planets are valuable objects for our understanding of planetary formation and evolution, with further discoveries needed to provide observational constraints on these interesting and complex systems. Their signal may be present in datasets from other transit surveys such as WASP \citep{Pollacco:2006gb} or NGTS \citep{Wheatley:2013uw}. Detecting these planets via the transit method presents observational challenges, as they exhibit transit timing variations (TTVs) on the order of days in magnitude, in addition to changes in the shape and duration of transits.

The purpose of this paper is to present constraints on the observational characteristics of a transiting circumbinary exoplanet through our knowledge of the host binary system, using a fast method which requires no complex modelling. In this way we aim to aid detection through reducing the problems generated by the large scale TTVs mentioned above. Specifically we address TTVs in coplanar circumbinary systems, placing general limits on the magnitude of such variations, through constraining the location of possible transits. A similar analysis was carried out for the system KIC002856960 \citep{Armstrong:2012ie,Lee:2013ee}, which shows similar large scale TTVs, and multiple transits per orbit, albeit in a triple star scenario. These constraints are of use to surveys for such planets, where we can place limits on and aid the design of new automated searches, such as the QATS algorithm \citep{Carter:2013bg}. While it is possible with numerical simulations to predict exact times of transit for circumbinary systems, our analytical model allows (under some approximations) constraints to be placed on systems where some or many orbital parameters are not yet known, including the majority of eclipsing binaries in the Kepler Eclipsing Binary Catalogue \citep{Prsa:2011dx,Slawson:2011fg}.

TTVs on the transits of circumbinary planets have two main sources. The first is a geometrical timing variation (we refer to this as Effect I) resulting from the changing positions of the host binary stars. This leads to a range in time in which transits can occur, similar to more `usual' TTVs, and is derived in Section \ref{sectGTVderiv}. The second is a precessional variation (referred to as Effect II), a long term oscillation in time around a constant periodicity of the potential location of transits, caused by precession of the planet's orbit (which is itself caused by torques arising from the non point mass nature of the binary). It is treated in Section \ref{sectPTVderiv}. There are other possible sources of TTVs, such as another planet in the system. The effect of such a planet, or any other known source of TTVs, is negligible compared to the above in circumbinary systems (c.f. Kepler-47b,c where planet-planet interactions are negligible \citep{Orosz:2012ku}).

We make use of several unusual terms in this paper, and define them here for clarity. First, a `crossing', or `crossing region'. This is the region of a circumbinary exoplanet's orbit where the planet crosses the binary star orbit, from the observer's perspective. It may only transit the stars within this crossing region, but will generally spend most of its time in the region out of transit. Second we use extensively the `azimuthal' period of a circumbinary planet, mentioned in LL. There are several periods which may be relevant to a circumbinary planet, and we make use of two here - the azimuthal period and the Keplerian period. The azimuthal period is the period which on average the planet takes between successive alignments with the observer, i.e. to traverse $2\pi$ radians relative to a fixed reference vector and plane. The Keplerian period is an osculating period taken at a particular epoch, derivable from Kepler's third law via the binary mass and planet semi-major axis. These two periods are not equivalent, and are discussed further in Section \ref{sectdiscuss}.
  
The structure of the paper is as follows. Section \ref{sectkepapprox} describes a Keplerian approximation which can be used to estimate the possible location of transits for a general planet and binary. Section \ref{sectnummod} describes the implementation of a numerical model used to test this approximation, with application to a demonstration simulated system. Section \ref{sectresults} shows the results of testing the analytical model against a distribution of binaries and planets modelled numerically, and applies the models to Kepler-16b, -34b and -35b. Section \ref{sectdiscuss} discusses the accuracy and usefulness of these results, as well as discussing observational issues in the search for circumbinary transiting exoplanets.

\section{Models}
\subsection{Analytic Approximation}
\label{sectkepapprox}

We present here a derivation which allows the potential location of transits of a circumbinary planet to be estimated without the need for detailed modelling or any free parameters. It proceeds using Keplerian orbital equations for both the stars and planets of a circumbinary system, and hence is an approximation only, as it does not consider three-body effects that perturb the orbits of the binary and planet (although precession of the planet's argument of periapse is included). We consider the TTVs of transits of only one star at a time, through this paper star 1. To consider transits of star 2, swap the indices 1 and 2 in Equation \ref{betaeqn}.

\subsubsection{Geometrical Timing Variations - Effect I}
\label{sectGTVderiv}
These variations arise from the movement of the binary stars within their orbit. As such we use the limits of this orbit, coupled with the time the planet takes to cross said orbit. We make use of an equation for the duration of a transit in a single star/planet system (Equation \ref{durationeqn}, from \citet{exoseagerdur}, their Equation 14). A crossing (defined in Section \ref{sectintro}) of a circumbinary planet is analogous to the transit of a single star by a planet passing in front of it; conceptually, we just replace the single star with a `metastar' of diameter equal to the maximum extent of the binary's orbit, giving

\begin{equation}
\label{durationeqn}
 T_{\textrm{GTV}} =  \frac{P_\textrm{p}}{\pi}\arcsin\left(\frac{R_{\textrm{metastar}}}{a_\textrm{p}}\right) \frac{\sqrt{1-e_\textrm{p}^2}}{1+e_\textrm{p}\sin(\omega_\textrm{p})},
\end{equation}
where $ T_{\textrm{GTV}}$ represents the duration of the crossing, subscript p represents the planet, $P$ the azimuthal period, $a$ the semi-major axis, $e$ the eccentricity and $\omega$ the argument of periapse. We have made the approximation that the impact parameter $b_\textrm{p}\ll R_{\textrm{metastar}}$, the inclination of the planet $i_\textrm{p} = \pi/2$ and $R_\textrm{p}\ll R_{\textrm{metastar}}$. To find $R_{\textrm{metastar}}$ we must derive the extent of the binary's orbit, projected onto the sky.

Consider the eclipsing binary orbit to be in the x-z plane, with the z axis being along the line of site of the observer. By doing this we take the binary orbit to have inclination $\pi/2$, a reasonable approximation for detached eclipsing binaries and for this purpose. Take the motion of star 1 in the x plane, projected onto the sky. From \citet{exoseagerskyproj} (their Equation 53, with $\Omega = 0$), this is given by

\begin{equation}
\label{skyprojx}
X = \beta(f)a_\textrm{b},
\end{equation}
where
\begin{equation}
\label{betaeqn}
\beta(f_\textrm{b}) = \frac{M_2}{M_1+M_2}\frac{(1-e_\textrm{b}^2)}{1+e_\textrm{b}\cos(f_\textrm{b})}\cos(\omega_\textrm{b}+f_\textrm{b}),
\end{equation}
and subscript b represents the binary, $f$ the true anomaly and $M_{1,2}$ the mass of stars 1 and 2 respectively. Taking the zero points of the differential with respect to $f_\textrm{b}$ of Equation \ref{skyprojx} gives us the minimum and maximum values of $X$ - the extent of the star's motion projected onto the sky. The values of the true anomaly of the binary at these points are given by


\begin{equation}
\label{minmaxf}
f_0,f_1 = \arcsin[-e_\textrm{b}\sin(\omega_\textrm{b})] - \omega_\textrm{b}.
\end{equation}

Equation \ref{minmaxf} has two solutions within the range $0,2\pi$. Inserting both into Equation \ref{skyprojx} gives the maximum and minimum values for $X$. We term these $X_1$ and $X_0$. Which of $X_0$ and $X_1$ is the minimum and which the maximum depends on $\omega_\textrm{b}$, but is unimportant here.

The radius of the `metastar' is given by

\begin{equation}
\label{rmetstareqn}
R_{\textrm{metastar}} = \frac{\vert X_1\vert + \vert X_0\vert}{2},
\end{equation}
and a scaled radius by
\begin{equation}
\label{rscaledeqn}
R_{\textrm{m,scaled}} = \frac{R_{\textrm{metastar}}}{a_\textrm{b}} = \frac{\vert \beta(f_1)\vert + \vert \beta(f_0)\vert}{2}.
\end{equation}
Substituting Equation \ref{rmetstareqn} into Equation \ref{durationeqn} leads to

\begin{equation}
\label{phaserangeeqn}
T_{\textrm{GTV}} = \frac{P_\textrm{p}}{\pi}\arcsin\left[R_{\textrm{m,scaled}}\left(\frac{P_\textrm{b}}{P_\textrm{p}}\right)^{\frac{2}{3}}\right] \frac{\sqrt{1-e_\textrm{p}^2}}{1+e_\textrm{p}\sin(\omega_\textrm{p})},
\end{equation}
where the ratio of semi-major axes has been substituted to the equivalent ratio of periods using Kepler's third law, allowing the use of the azimuthal period outlined in Section \ref{sectintro}. In the presented form $ T_{\textrm{GTV}}$ represents the duration of a crossing, and as such a range of time within which transits can occur. The argument of periapse, $\omega_\textrm{p}$, is a function of time due to precession of the planetary orbit; assuming a constant precession rate it can be estimated analytically using Equation 5 of \citet{Doolin:2011ib}, hereafter DB, which is derived from that of \citet{Farago:2010ev}.

Lacking knowledge of the present system alignment, it is possible to take a `safe' approximation by using the value of $\omega_\textrm{p}$ which gives the maximum $ T_{\textrm{GTV}}$, i.e. $\omega_\textrm{p} = 3\pi/2$. This corresponds to when the planet transits near its apoapse, and hence is travelling relatively slowly so that the range of transit times is extended. Using this constant value of $ T_{GTV}$ is often more practical. For systems with low planetary eccentricity the variation caused by varying $\omega_\textrm{p}$ is small (on the order of a few percent in $ T_{GTV}$).

\subsubsection{Precessional Timing Variation - Effect II}
\label{sectPTVderiv}
This variation is caused by the precession of the planet's orbit. For an eccentric planetary orbit, this precession will result in shifts in the time of potential transits away from the `expected' time for a constant periodic signal. The magnitude of these shifts at a given time depends on the instantaneous value of $\omega_\textrm{p}$.

We assume a constant precession rate for the planetary orbit, such that

\begin{equation}
\label{eqnomrate}
\frac{d \omega_\textrm{p}}{dt} = \frac{2\pi}{P_{\omega}},
\end{equation}
where $P_{\omega}$ represents the period of precession of the planet's periapse, and can be estimated analytically through the equation of DB.

For a planet precessing in the prograde direction, this change in $\omega_\textrm{p}$ represents time `gained', a portion of its orbit which it does not have to cover before aligning with the observer once more. The differential amount of time saved (i.e. period shifted) in this way is given by

\begin{equation}
\label{eqntimeloss}
\frac{dP_\textrm{p}}{d \omega_\textrm{p}} = \frac{dt}{df_\textrm{p}},
\end{equation}
where $dP$ represents an apparent change in the period of the planet, and $f_\textrm{p}$ is the true anomaly of the planet, with $dt/df_\textrm{p}$ evaluated at $f_\textrm{p}=\pi/2 - \omega_\textrm{p}$, the value of $f_\textrm{p}$ at transit conjunction.

There are two contributions here, a constant term from the precession and a varying oscillation induced by the effect of the eccentricity of the planet's orbit. The constant term can be found simply, by realising that the planet `loses' one full orbit of time in one precessional period. For a constant precession rate, this gives a constant rate of time loss of $P_\textrm{p}/P_{\omega}$, which must be subtracted from Equation \ref{eqntimeloss} to find the oscillation term. When using the azimuthal period of the planet (as defined in Section \ref{sectintro}), or searching observationally for transits this constant term is automatically accounted for, which is why it must be removed here.

Continuing the derivation, we take the standard Keplerian orbital equation for $df_\textrm{p}/dt$ \citep[their Equation 32]{exoseagerdfbydt} evaluated at $f_\textrm{p} = \pi/2 - \omega_\textrm{p}$,
\begin{equation}
\label{eqndfbydt}
     \frac{df_\textrm{p}}{dt} = \frac{2\pi}{P_\textrm{p}} \frac{[1+e_\textrm{p}\sin(\omega_\textrm{p})]^2}{(1-e_\textrm{p}^2)^{\frac{3}{2}}},
\end{equation}
where we have approximated $P_\textrm{p} \simeq P_\textrm{p}(1+1/P_{\omega})$. Combining Equations \ref{eqnomrate} and \ref{eqntimeloss} gives us the oscillation term

\begin{equation}
 T_{\textrm{PTV}} = \int_{t_0}^{t}dP_\textrm{p} = \int_{t_0}^{t}\left(\frac{dt}{df_\textrm{p}}  - \frac{P_\textrm{p}}{P_{\omega}}\right) d \omega_\textrm{p},
\end{equation}
which, after inserting Equation \ref{eqndfbydt}, becomes
\begin{equation}
\label{eqnPTV}
 T_{\textrm{PTV}} = -\frac{P_\textrm{b}}{P_{\omega}} \int_{t_0}^{t} \left(\frac{(1-e_\textrm{p}^2)^{\frac{3}{2}}}{(1+e_\textrm{p}\sin[\omega_\textrm{p}(t)])^2} - 1\right) dt,
\end{equation}
where the negative sign accounts that this is time gained or equivalently an apparent shortening of the planetary period, and applies for prograde precession. The quantity $ T_{\textrm{PTV}}$ represents an oscillation of the location of possible transits with time. We give an example of its effect through application to a demonstration simulated system in Section \ref{sectnummod}.

\subsubsection{Combined TTV Limits - Practical Use}
\label{sectcombinedlimits}
Equations \ref{phaserangeeqn} and \ref{eqnPTV} can be combined to provide limits on the TTVs of transiting coplanar circumbinary planets. At a given epoch, $ T_{\textrm{PTV}}$ represents the offset around some zero point that the range of possible transit times would be centred around, whereas $ T_{\textrm{GTV}}$ represents the extent of the range around this offset. We present constraints here for practical use, in the situation where one or more transits have been detected, and limits need placing on the times of as yet undetected transits. A period must be estimated, either from the separation of two transits (or fractions of this) or by using a succession of trial periods. In the case where only one or two transits are known, as we do not know where in the possible transit range the transit falls we must use double the range to cover all possible times, giving the following limits:
\begin{equation}
\label{eqntmin}
t_{\textrm{min}} (i) = t_0 + iP_\textrm{p} + T_{\textrm{PTV}}(t_0 + iP_\textrm{p}) - T_{\textrm{GTV}}(t_0 + iP_\textrm{p})
\end{equation}
and
\begin{equation}
\label{eqntmax}
t_{\textrm{max}} (i) = t_0 + iP_\textrm{p} + T_{\textrm{PTV}}(t_0 + iP_\textrm{p}) + T_{\textrm{GTV}}(t_0 + iP_\textrm{p}),
\end{equation}
where $t_0$ represents the time of first transit, and $i$ an index for the orbit under consideration (each orbit may contain more than one transit, though in practice this is unusual). The quantities $t_{\textrm{min}}$ and $t_{\textrm{max}} $ represent the minimum and maximum times between which possible undetected transits must fall within on each orbit, for the case of one or two known transits.

\begin{figure*}
\resizebox{\hsize}{!}{\includegraphics{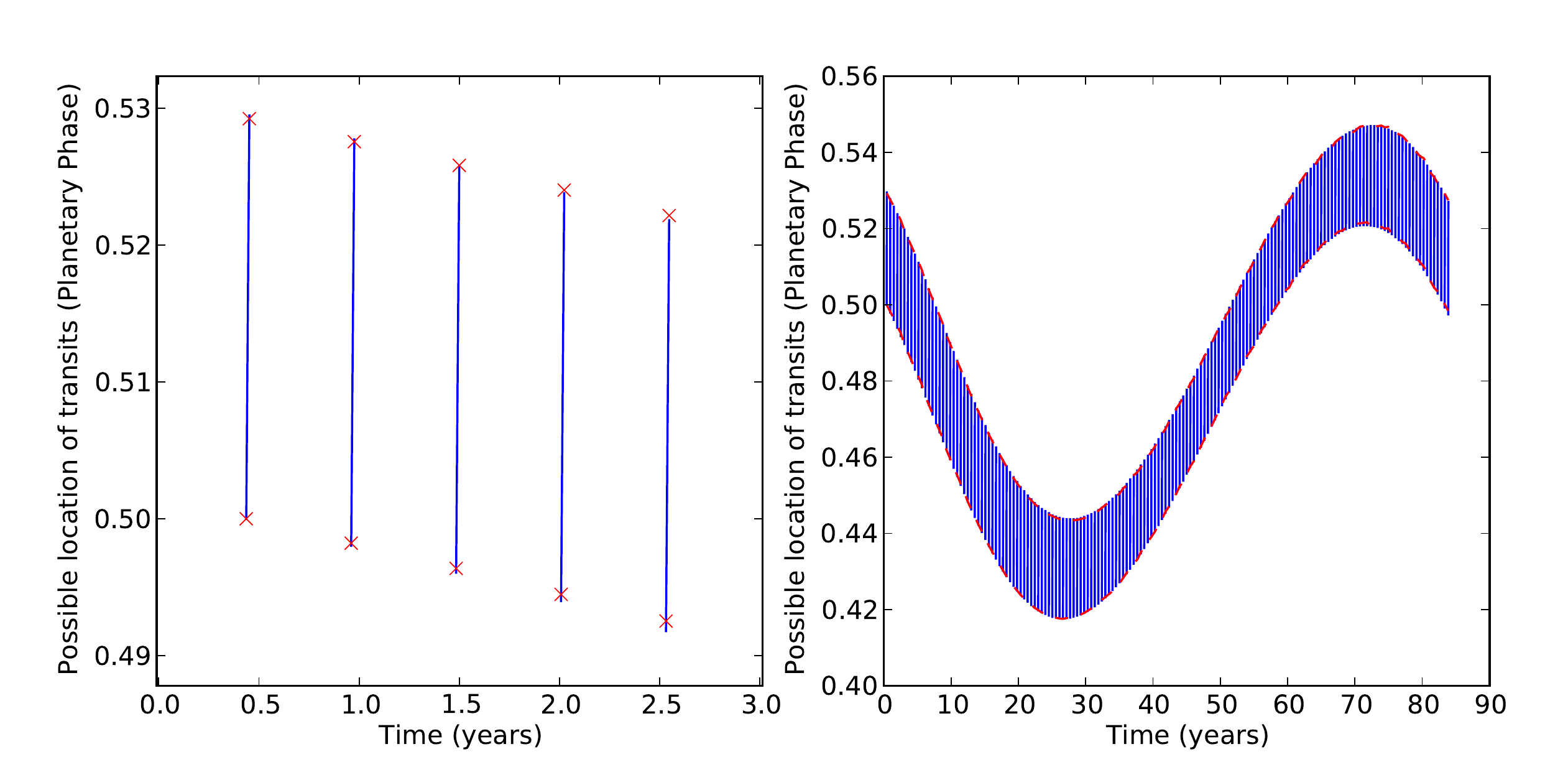}}
\caption{\textbf{Left} Three years and five crossings of a simulated planet. Transits must occur on the lines. The crosses represent the predicted maximum and minimum time for each crossing region derived from our analytical equations. The length in phase of each line represents $T_{\textrm{GTV}}$, while the shifting of the lines in phase represents $T_{\textrm{PTV}}$. The phase is calculated through phase folding over the planetary azimuthal period (191.5 days). The starting epoch $t=0$ is arbitrary, as are absolute values of the phase. \textbf{Right} As left for a full planetary precessional period. The dashed line shows the analytical equation prediction, realigned with the numerical model every three years (chosen as a representative length for an observing campaign.) Realignment is justified as this is how the equations would be used in practice, with a single detected event representing a zero point to which the equations would be aligned.}
\label{t1t2demo}
\end{figure*}

Over short ($\ll P_{\omega}$) timescales $T_{\textrm{GTV}}$ is the dominant contribution (in some systems, such as those with low eccentricity planets, it is always so), and $T_{\textrm{PTV}}$ may be neglected. Using the maximum possible value of $T_{\textrm{GTV}}$ (by setting $\omega_\textrm{p}=3\pi/2$ in Equation \ref{phaserangeeqn}) provides a `safe' (in that the result will always be an overestimate) way of neglecting the time and $\omega_\textrm{p}$ dependence of $T_{\textrm{GTV}}$. Similarly, if little is known about a proposed circumbinary system, parameters in the above equations can be easily approximated with only small and quantifiable errors introduced.

The effects of $T_{\textrm{GTV}}$ and $T_{\textrm{PTV}}$ are shown for a demonstration circumbinary planet in Section \ref{sectdemo}.
 
\subsection{Numerical Model}
\label{sectnummod}
\subsubsection{Approach}
We use a numerical model to test the above analytical framework. The N-body equations of motion were integrated using a fourth-order Runge-Kutta algorithm. Since this integrator does not inherently conserve energy, the total system energy was calculated over time to ensure that it was conserved such that the energy loss fraction remained below approximately $10^{-7}$.

To calculate the azimuthal period numerically we averaged the time intervals between the planet passing each of the two boundaries of the projected star orbit. The azimuthal period is the mean of these two averages. An alternative method is to average the interval between system centre of mass crossing times, which will converge to the same value but more slowly because it is only based on one crossing point, not two. Over time, the average interval between consecutive transits will converge to the azimuthal period.

\subsubsection{Demonstration}
\label{sectdemo}
We here apply the numerical and analytical models to a simulated system (chosen from the simulations of Section \ref{sectresults} as a system with a typical error) to demonstrate the effects of the derived timing variations. This system has a binary star with period 14.1 days, eccentricity 0.13, stellar masses of 1.22 and 1.07 $M_\odot$ and argument of periapse 282.3 deg. The planet has azimuthal period 191.5 days and eccentricity 0.16, leading to a precessional period for the planet of 84.2 years from the numerical model. Figure \ref{t1t2demo} shows the potential locations of planetary transits derived from the numerical model, using times of potential transit phase-folded at the above azimuthal period. Potential transits must occur on each solid line. The variations seen are discussed in Sections \ref{T1var} and \ref{T2var}.

\subsubsection{Geometrical Timing Variations - Effect I}
\label{T1var}
The Effect I geometrical timing variations introduced in Section \ref{sectintro} and derived in Section \ref{sectGTVderiv} arise from the significant motion of the host binary stars. The planet can take several days to traverse the full extent of the binary orbit, and it is during this time that transits will occur. The TTVs, given by Equation \ref{phaserangeeqn}, can therefore be very large. By considering circular orbits and Solar-mass stars, Equation \ref{phaserangeeqn} can be approximated by $T_{\textrm{GTV}} \approx (P_\textrm{p} P_\textrm{b}^2)^{1/3}/(2\pi)$, with periods in days, which demonstrates the size of the TTVs and their period dependence.

This geometrical contribution to the TTVs corresponds to the length of the lines in Figure \ref{t1t2demo}. The magnitude of the Effect I term itself oscillates with the precession period of the planet, due to the changing speed of the planet at crossing, as different regions of its eccentric orbit line up with the observer.

\subsubsection{Precessional Timing Variations - Effect II}
\label{T2var}
The other variation, an oscillation in phase or equivalently oscillation in apparent period, is due to the precession of the planet causing transits to correspond to different phases, as seen in Figure \ref{t1t2demo}(right). The oscillation is in particular caused by the changing instantaneous effect of the precession on a planet in an eccentric orbit. This is different to the contribution of precession in the Effect I geometrical case, which varies $T_{\textrm{GTV}}$ due to the changing planetary velocity. The Effect II precessional variation becomes significant over timescales approaching the planetary precessional period, typically decades. The amplitude of this variation is strongly dependent upon the planetary precessional period and eccentricity.

\section{Results}
\label{sectresults}
\subsection{Setup}
The accuracy of the model of Section \ref{sectkepapprox} was tested using the numerical model (Section \ref{sectnummod}) applied to a simulated distribution of 1\,000 single-planet circumbinary systems, 799 of which were stable over 1\,200 years (longer than the maximum planetary precession period found, and significantly longer than the majority). A more thorough stability analysis was not deemed necessary for the purposes of testing the equations in this paper. The binary star periods and eccentricities were taken from \citet{Halbwachs:2003}, which presented an unbiased distribution taken from radial velocity surveys, expanding upon the work of \citet{Duquennoy:1991wk}. The primary star masses were taken from the Kepler catalog of all stars monitored, using an empirical calibration from Torres \citep{Torres:2010eoa} to calculate the mass based on the metallicity, effective temperature and log $g$. The secondary star mass was determined using the mass ratio distributions found in \citet{Halbwachs:2003}, for binaries with periods less than and greater than 50 days. The radii of the stars were unimportant for this test.

For the planets, since no circumbinary planet distribution is known as yet, the period and eccentricity distributions were taken from data for planets orbiting single stars. Only radial velocity data were used to avoid the bias towards small periods seen in transit surveys. The planet was taken as a massless test particle, as its mass has a minimal effect on the dynamics. The planet radius was also unimportant for this simulation, as it has no effect on the dynamics. For each circumbinary system the minimum planet period was four times that of the binary, as a rough stability constraint \citep{Holman:1999aa}, although some systems still proved to be unstable (particularly those with high eccentricities). The maximum planet period was set at 500 days, long enough that TTVs in such systems are unlikely to be of interest in the near future. All systems were exactly coplanar. Each of these systems was integrated numerically over its expected precession period (calculated from the equation of DB) with a time step of 30 minutes. The system's azimuthal period was then calculated from the time it took the planet to orbit the system centre of mass on average.

To test the analytical model we used Equations \ref{phaserangeeqn} and \ref{eqnPTV} to predict the limits on possible transit time of the simulated planets. The precession period was split into three-year baselines (chosen as the length of a representative observing campaign). At each of the three-year baseline for each system, the predicted and numerical limits were initially aligned (as would be the case when detecting the first transit of a candidate planet) and then the system and predicted limits were allowed to evolve. At each crossing, the deviations between the upper analytical and numerical limits and lower analytical and numerical limits were averaged, and the same averaged for all crossings within each of the three-year baselines.

\subsection{Test Results}
\begin{figure}
\resizebox{\hsize}{!}{\includegraphics{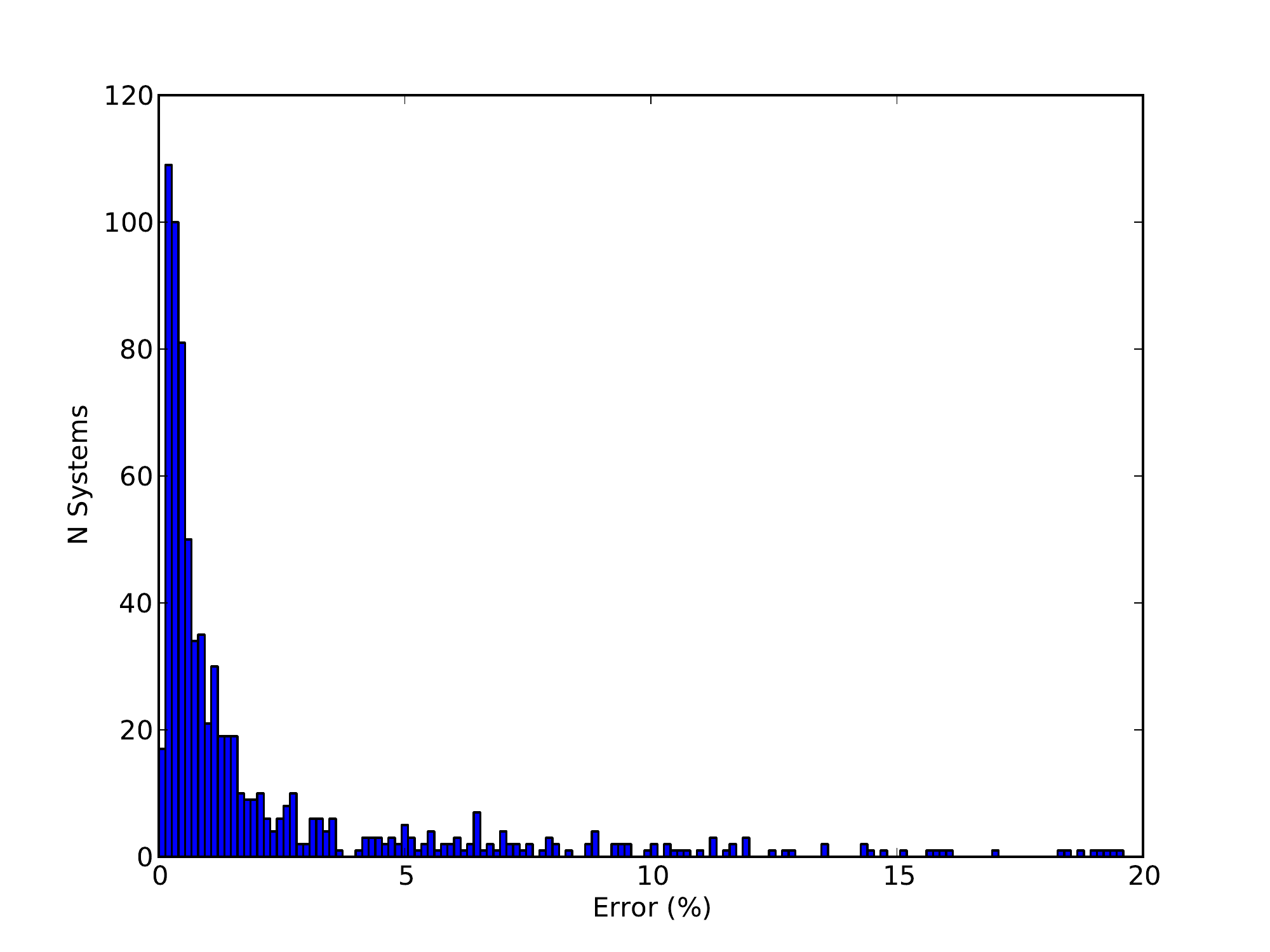}}
\caption{The error in comparing simulated numerical limits on the possible transit locations of 756 systems to the predictions of combining Equations \ref{phaserangeeqn} and \ref{eqnPTV}. The difference between the analytical and numerical models is expressed as a percentage of the numerical planet crossing duration at each crossing. For clarity, 43 additional systems with errors greater than 20\% are not shown for clarity. Four of these systems have errors over 100\%.}
\label{figtest}
\end{figure}

Results are represented as a percentage of the numerically integrated crossing time found at each planetary crossing. As such, an error of 100\% represents analytically predicted transit limits which are misaligned by one crossing time on average. Figure \ref{figtest} shows the histogram of percentage errors found for the 799 stable systems. The peak shows an error of 0.4\%. The median error is 0.84\%. For clarity, 43 systems are not shown in Figure \ref{figtest}. These represent badly predicted single systems, with percentage errors higher than 20\% (four of them have errors over 100\%). These larger error systems are discussed in Section \ref{sectdiscuss}.

\subsection{Application to Kepler-16b, -34b and -35b}
\begin{figure}
\resizebox{\hsize}{!}{\includegraphics{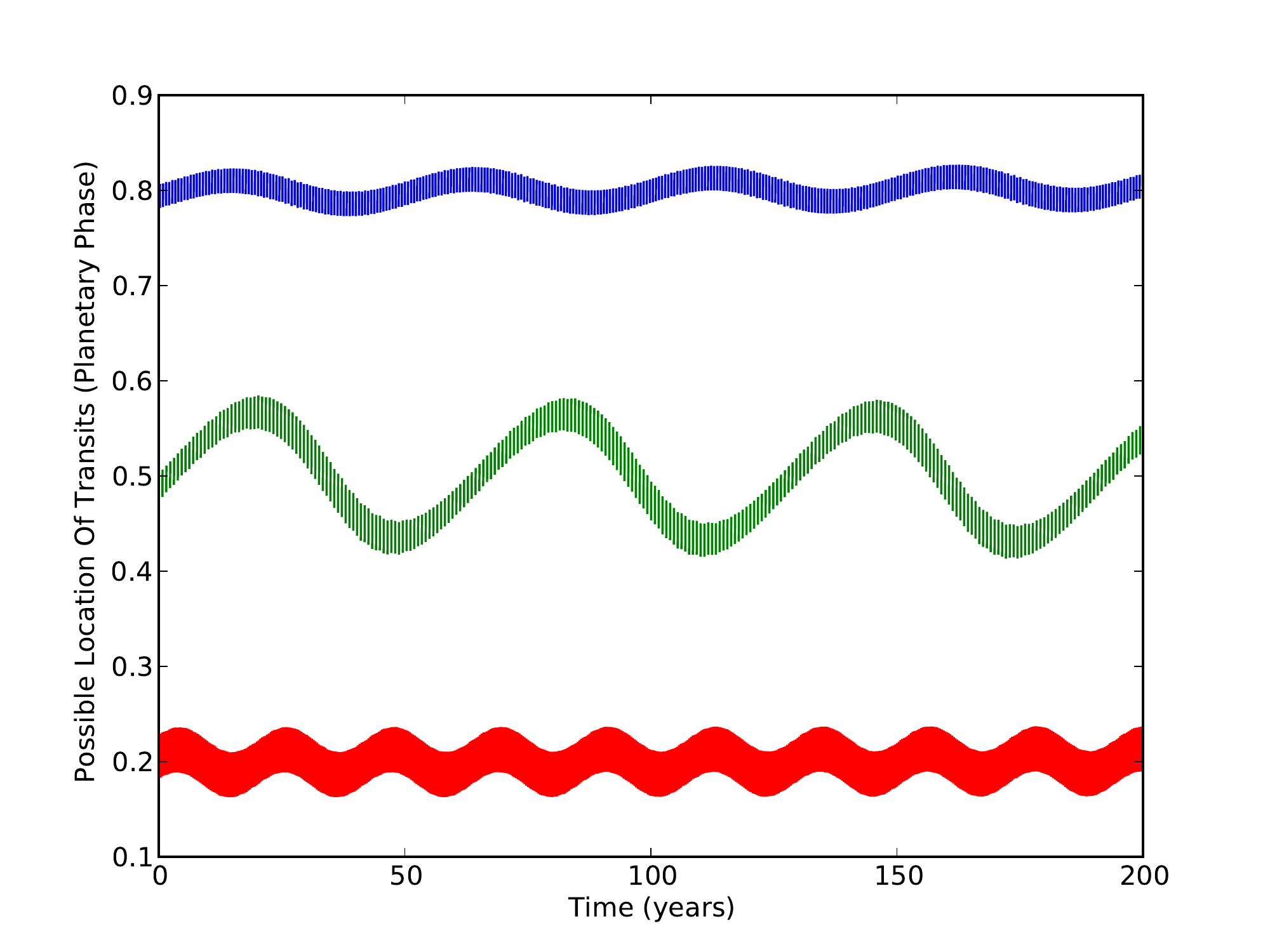}}
\caption{The variation in planetary phase of potential transit times, derived from the numerical model. Transits must occur within the thick bands. From top to bottom, the lines show Kepler-16b, -34b and -35b. The phase is calculated using the planetary azimuthal period in each case. Absolute values of the phase are arbitrary. The starting epoch $t=0$ is also arbitrary.}
\label{phasevar}
\end{figure}

The numerical model was applied to the known systems Kepler-16b, -34b and -35b, and times of possible transit were extracted. We find azimuthal periods of 227.06, 283.13 and 127.30 days for -16b, -34b and -35b, respectively. These are slightly offset from those found by LL. These are compared to Keplerian periods from the respective discovery papers of 228.78, 288.82 and 131.46 days \citep{Doyle:2011ev,Welsh:2012kl}. We note that care must be taken regarding the different reference frames parameters for these planets can be published under, and also regarding the instantaneous and highly variable nature of many of the usual planetary parameters. Figure \ref{phasevar} shows the potential locations of planetary transits derived from the numerical model, using times of potential transit phase-folded at the above azimuthal periods. Potential transits must occur within the thick band for each planet. The thickness of each band represents the Effect I, geometrical timing variation, and the oscillation in phase of the band represents the Effect II, precessional variation. The amplitude of this Effect II variation is strongly dependent upon the planetary precessional period and eccentricity. The period of the Effect II oscillations is equivalent to the planet's precessional period, {\raise.17ex\hbox{$\scriptstyle\sim$}}48, {\raise.17ex\hbox{$\scriptstyle\sim$}}63 and {\raise.17ex\hbox{$\scriptstyle\sim$}}21 years for Kepler-16b, -34b and -35b, respectively.\\

Whilst the previous large-scale test used massless test particles, in this application the planet masses were included in the N-body code. To demonstrate that the planet mass has only a small effect on the transit times, we also simulated the Kepler planets with zero mass. The transit times of the mass and massless simulations were compared over a 200 year period in blocks of three years. On average the difference, as a percentage of the TTV range, was only 3.6\%, 0.8\% and 0.7\% for Kepler-16b, -34b and -35b, respectively. Future work may include updated equations that incorporate the planet mass, but this will likely only be beneficial for very massive circumbinary planets.

\begin{figure}
\resizebox{\hsize}{!}{\includegraphics{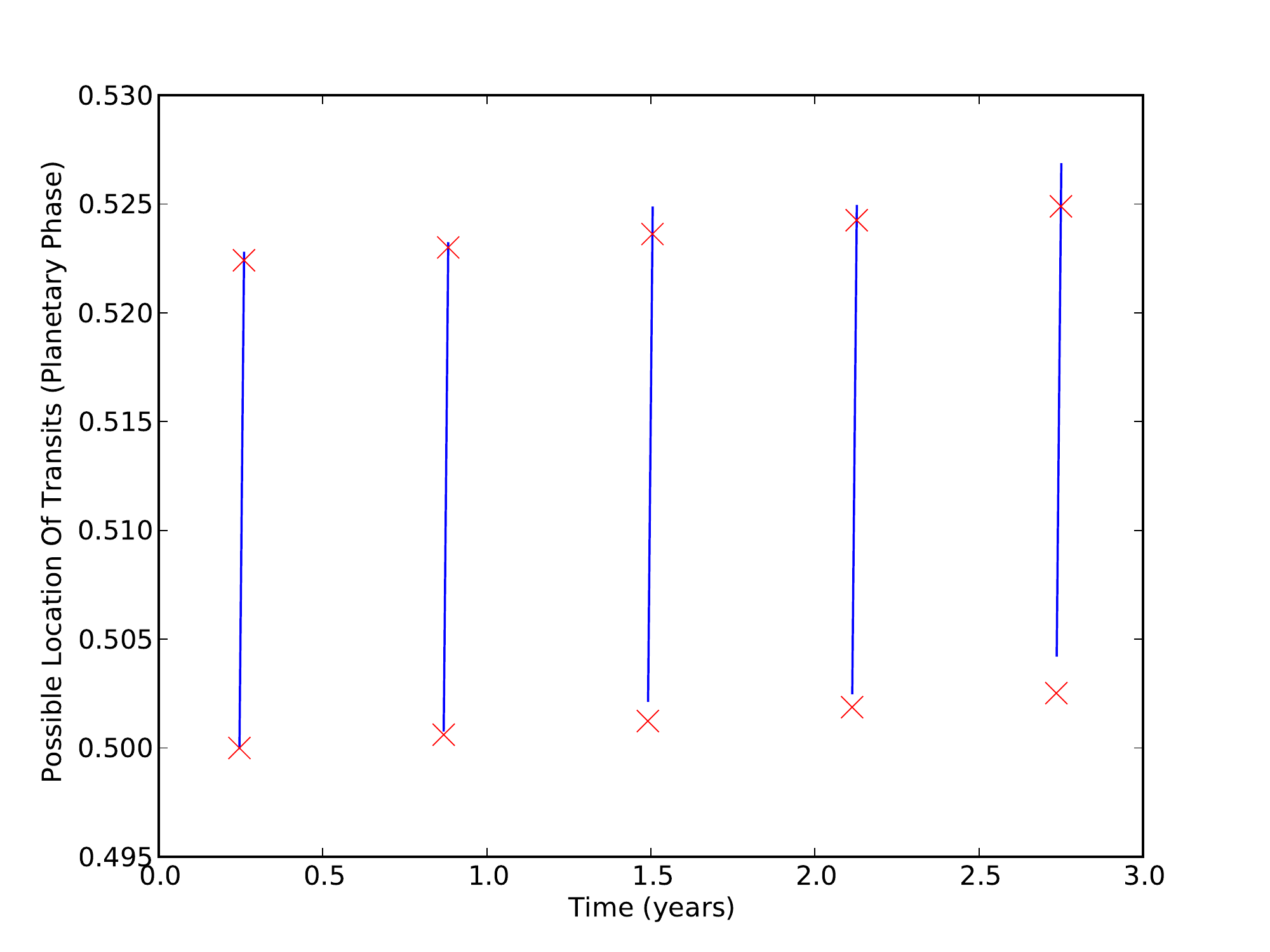}}
\caption{A typical three-year region of the Kepler-16b curve of Figure \ref{phasevar}. Transits must occur on the lines. The crosses represent the predicted maximum and minimum time for each crossing region derived from our analytical equations.  The length in phase of each line represents $T_{\textrm{GTV}}$, while the shifting of the lines in phase represents $T_{\textrm{GTV}}$. Absolute values of the phase are arbitrary. The starting epoch $t=0$ is also arbitrary.}
\label{16dbtest}
\end{figure}
\begin{figure}
\resizebox{\hsize}{!}{\includegraphics{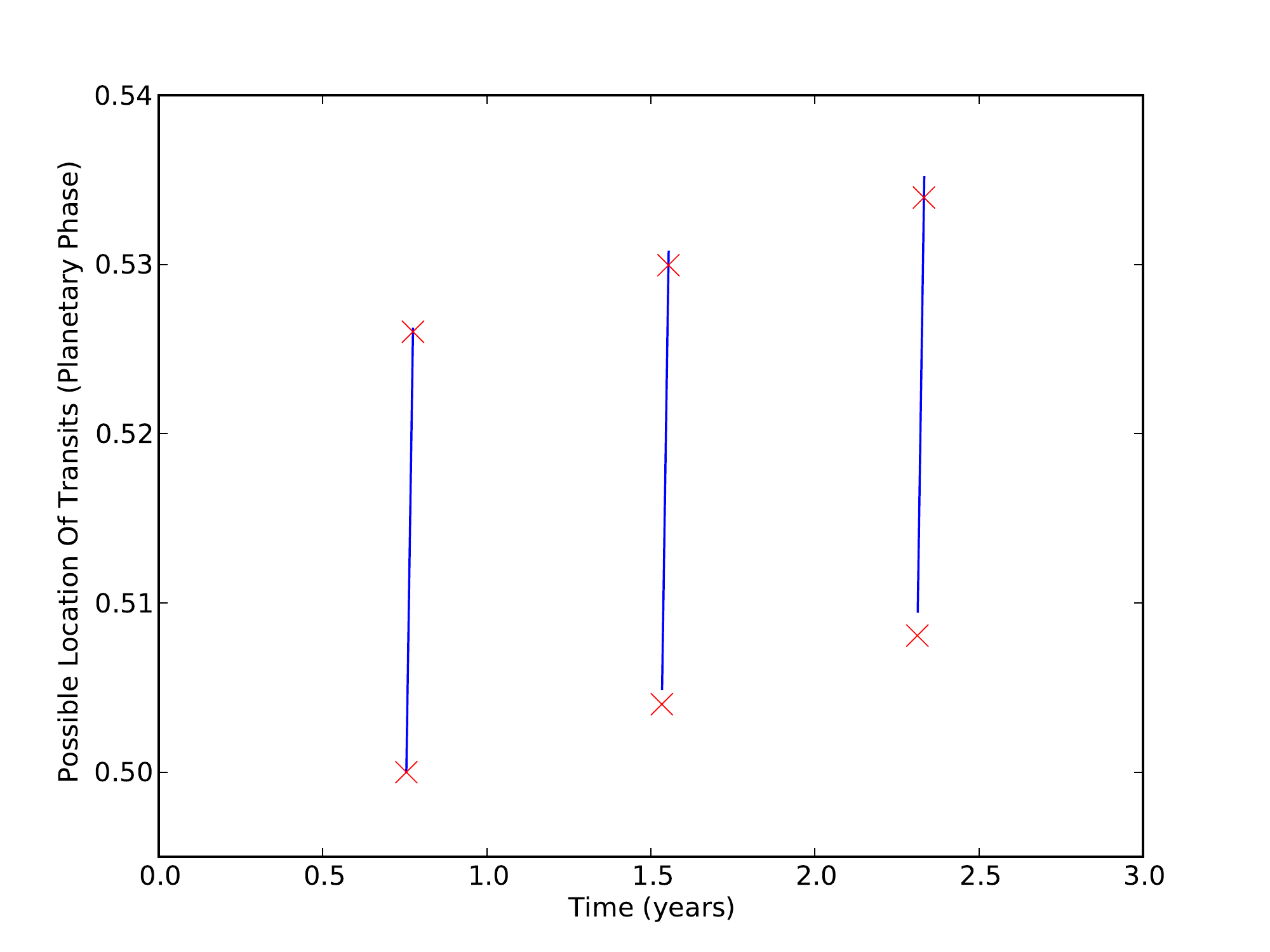}}
\caption{As Figure \ref{16dbtest} for Kepler-34b.}
\label{34dbtest}
\end{figure}
\begin{figure}
\resizebox{\hsize}{!}{\includegraphics{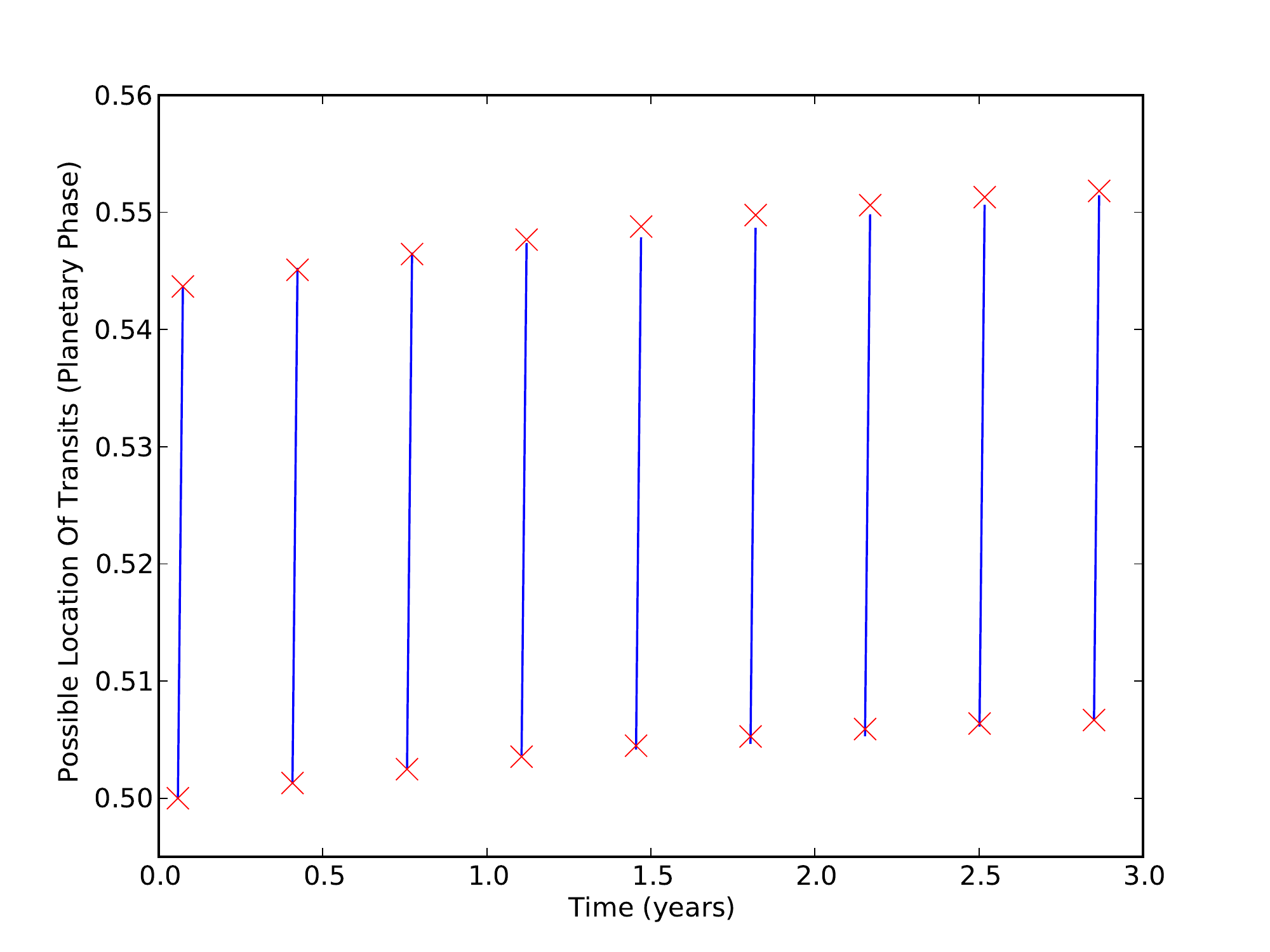}}
\caption{As Figure \ref{16dbtest} for Kepler-35b.}
\label{35dbtest}
\end{figure}

A typical three-year region is shown for each planet in Figures \ref{16dbtest}, \ref{34dbtest} and \ref{35dbtest}, with the analytical model prediction for each crossing. We note the slight secondary oscillation in Figure \ref{16dbtest}. This is an additional dynamical effect likely due to a non-Keplerian effect of the host binary, and is stronger for Kepler-16b than for -34b or -35b. We do not attempt to predict this effect in this work.

\section{Discussion}
\label{sectdiscuss}
\subsection{Overview}
We have derived and validated a fast and simple to implement framework for placing limits on the possible locations of transits for a transiting coplanar circumbinary exoplanet. These variations can be split into two parts - Effect I, geometrical, caused by the changing positions of the binary stars as they orbit, and Effect II, precessional, caused by the long term precession of the planet's orbit. With the likelihood of future searches for circumbinary exoplanets high, and the possibility of discovering such planets in the extant data from previous surveys, being able to place limits on the location in time of potential signals is particularly useful.

\subsection{Accuracy}
\label{sectdiscussacc}
Figure \ref{figtest} shows the accuracy of Equations \ref{phaserangeeqn} and \ref{eqnPTV} in predicting the possible times of transit of coplanar circumbinary planets - a median percentage error of 0.84\% of the planet crossing time across the test set of 799 stable systems, over three years of observations. This can be used as an error when using Equations \ref{eqntmin} and \ref{eqntmax} to predict possible times of transit, where the percentage error should be applied to both $t_\textrm{max}$ and $t_\textrm{min}$. We note that our stated errors depend on the time baseline covered - they will be reduced for baselines lower than three years, and increased for those higher. The stated errors should, however, be indicative for a general observing campaign. Limitations on the accuracy arise primarily from non-Keplerian effects (beyond simple constant precession of the planet's orbit, which we account for). This is demonstrated by the 43 systems with errors greater than 20\%, including four with errors greater than 100\%. These, and the scattered systems found at over 5\% in Figure \ref{figtest}, are systems which appear to be stable but which show strong dynamical effects we have not accounted for, such as shorter period additional oscillations of $\omega_\textrm{p}$ or other effects we do not investigate here. The underlying dynamics behind these are beyond the scope of this paper. Encouragingly, it seems that such effects are strong only in a small minority of cases - the analytical model missed the possible transit range entirely in only 0.5\% of the tested stable systems. 

\subsection{Applications}
We anticipate that Equations \ref{phaserangeeqn} and \ref{eqnPTV} (and in practice Equations \ref{eqntmin} and \ref{eqntmax}) will prove useful particularly for current and future searches for circumbinary planets. They provide a link between our theoretical knowledge of a circumbinary planetary system and the observational transit signatures which may arise from it, without requiring complex modelling or N-body integrations. This can be used to place limits on the potential transit times of candidate planets around a binary star, for the purpose of constraining searches for the transits of unknown planets. As a specific example, Equations \ref{eqntmin} and \ref{eqntmax} can be used to set the parameters $\triangle_{\textrm{min}}$ and $\triangle_{\textrm{max}}$ in the QATS search algorithm \citep{Carter:2013bg}. Importantly, this analytic framework can be used on systems where detailed knowledge of the component stars and orbital parameters is lacking, something impossible for N-body models. Full use of Equations \ref{eqntmin} and \ref{eqntmax} requires knowledge of the binary system, specifically the individual stellar masses, binary orbital eccentricity, argument of periapse and binary period, as well as the argument of periapse and eccentricity of the planet (while the planetary period is involved, we envisage that for general searches for unknown planets a series of trial periods would be used). Lacking some or all of these details, it is possible to make useful conclusions through using simplifying assumptions - taking $M_2\ll M_1$ for example removes the need for knowledge of the stellar masses while only overestimating the Effect I timing variation limit by at most a factor of two (i.e. placing loose but still useful limits on transit timing).

We repeat that it is possible in general to neglect the time and $\omega_\textrm{p}$ dependent part of $T_{\textrm{GTV}}$ (by setting $\omega_\textrm{p} = 3\pi/2$, as the value which gives the maximum value of $T_{\textrm{GTV}}$). It is also possible for low eccentricity planets to neglect $T_{\textrm{PTV}}$. This makes Equations \ref{eqntmin} and \ref{eqntmax} simply a constant limit on the TTV of a planet. These additional terms are however included in this work so that they can be utilised if necessary, particularly for highly eccentric planets or those whose precessional periods approach the baseline of observations used. Note that Equations \ref{eqntmin} and \ref{eqntmax} represent double the range of transit times predicted by Equations \ref{phaserangeeqn} and \ref{eqnPTV}, as it would not be known where in this range a first detected transit fell. 

These equations are also useful in reverse, for making first estimates of planet parameters using the observed transit variations of a newly discovered planet candidate. In this situation, the planet azimuthal period must be estimated using the mean transit interval. With this, the maximum observed transit timing variation around this period can be obtained. Neglecting $T_{\textrm{PTV}}$, this represents a lower limit on Equation \ref{phaserangeeqn}. In the situation where the binary period, eccentricity and argument of periapse are known through the binary light curve, this gives a constraint on a combination of the planet eccentricity, argument of periapse and the binary mass ratio. The geometrical contribution $T_{\textrm{GTV}}$ is only weakly dependent on the planet eccentricity for moderate eccentricities, so by setting $e_\textrm{p}=0$ an approximate lower limit can be found on the binary mass ratio (for $e_\textrm{p}=0.2$ this approximation has an error of at most {\raise.17ex\hbox{$\scriptstyle\sim$}}20\%, depending on the precise value of $\omega_\textrm{p}$). Conversely, if the lower limit on $T_{\textrm{GTV}}$ found from the observed transits is especially high for the known binary parameters, this is an indication of high planetary eccentricity. Such constraints can be of use when attempting to find best fitting orbital solutions for these systems.

In the situation where for example only a few transits are detected, and the orbital solution is degenerate or poorly constrained (such that N-body integration is unfeasible), these expressions can be used for placing limits on the time period for which an object should be surveyed from the ground to detect future transits. This makes such follow up work much more efficient, and becomes relevant when continuous space based observations are not available.

\subsection{Observational Considerations}
We summarise here some issues which have become apparent affecting observational searches for circumbinary exoplanets. While this paper aims to reduce the difficulty caused by TTVs, these other limitations to detection of circumbinary planets remain and should be noted:
\begin{description}
\item
\textbf{Azimuthal Period} 
This is the time which on average the planet takes to traverse $2\pi$ radians in a fixed reference frame - i.e. the time interval between successive conjunctions. It is offset from, for example, the Keplerian period which can be derived from the planet's semi-major axis and the binary mass. In LL it is shown that the azimuthal period is shorter than the Keplerian orbital period for circumbinary planets. The effect of this can be seen in many of the published transiting circumbinary planets so far. If we take the observed times of transit of these planets and estimate a period from the mean transit interval (which is equivalent to the azimuthal period), the estimated period is generally found to be a few days under the published Keplerian period. This is not an error, but a mark of the difference between the azimuthal period and Keplerian period that LL mention. The effect is clear for Kepler-16b: The maximum TTV at the published Keplerian period (228.78 days) is {\raise.17ex\hbox{$\scriptstyle\sim$}}13 days, but at the azimuthal period we find (225.72 days), it is {\raise.17ex\hbox{$\scriptstyle\sim$}}4.5 days, significantly lower. This azimuthal period is the important quantity when considering circumbinary planets from an observational perspective.
\item
\textbf{Non-Coplanarity}
If a circumbinary planet is not close to coplanarity with its host binary (such that it is within a few degrees of the binary orbital plane), then due to the motion of the binary stars it will often `miss' them while crossing, exhibiting transits only on some orbits and again making detection much less likely. This constraint is relaxed for binary stars where the mass of one star is much greater than that of its companion (such that the more massive star's orbit is smaller than its radius) or for contact binaries. Furthermore, for systems that are not exactly coplanar, the precession of the planetary orbit will take it in and out of a transiting configuration. This is the case for Kepler-16, where the transits across the larger star A are predicted to cease in early 2018 and return in approximately 2042.
\item
\textbf{Eccentricity}
As part of the source of TTVs of circumbinary exoplanets is due to precession of the planet's orbit, highly eccentric planets will show more variations. While this does not reduce their detection chances as much as the above points, it increases the difficulty caused by these variations, further `blurring' the planet's transit signal. The `blurring' effect of eccentricity is then scaled by the period of the precession of the planet's orbit. Planets that precess faster will experience more transit timing variations over a given timescale.
\end{description}

\section{Conclusion}
\label{sectconc}
\begin{enumerate}
\item
There are two key contributions to the timing variations affecting transits of circumbinary planets. These are geometrical, Effect I, from the motion of the binary stars, and precessional, Effect II, from the precession of the planet's orbit. Other contributions, from for example other planets in the system, are generally on the order of minutes or less in amplitude and negligible compared to these.
\item
We have derived and validated an analytic framework to quickly estimate each of these terms, for a planet coplanar with its host binary.
\item
This can be used to place limits on the location of possible transits. In particular, the equations can be approximated using minimal knowledge of the system (in contrast to a more detailed numerical integrator), making them useful for searching datasets for transits of such planets or in reverse making first estimates of parameters using the observed transit variations. Specifically, full use of the equations require the individual stellar masses, binary eccentricity, argument of periapse and binary period, as well as the period, argument of periapse and eccentricity of the planet. It is simple to approximate the parameters or use trial values where necessary, as described at various points above.
\item
We have also summarised some observational issues which have become clear affecting the prospects of detection of circumbinary planets.
\end{enumerate}

\section*{Acknowledgements}
The authors are grateful for the useful comments of an anonymous referee, which helped improve the paper. D. V. Martin is funded by the Swiss National Science Foundation. A. H.M.J. Triaud is a Swiss National Science Foundation fellow under grant number PBGEP2-145594.

\bibliography{papers_17042013}
\bibliographystyle{mn2e_fix}

\end{document}